\documentclass{article}
\usepackage{amssymb,amsmath}

\newcommand{\A}{\hat{a}}
\newcommand{\Ad}{\hat{a}^\dagger}
\newcommand{\B}{\hat{b}}
\newcommand{\Bd}{\hat{b}^\dagger}
\newcommand{\dm}{\hat{\rho}}
\newcommand{\F}{\mathcal{F}}
\newcommand{\N}{\mathcal{N}}

\begin{document}

\title{Collision Integrals in the Kinetic Equations of dilute Bose-Einstein Condensates}

\author{Erich D. Gust and L. E. Reichl \\
The Center for Complex Quantum Systems,\\
The University of Texas at Austin, Austin, Texas 78712}
\date{\today }

\maketitle

\begin{abstract}
We derive the mean field kinetic equation for the momentum distribution of Bogoliubov excitations (bogolons) in a spatially uniform Bose-Einstein condensate (BEC), with a focus on the collision integrals. We use the method of Peletminksii and Yatsenko rather than the standard non-equilibrium Green's function formalism. This method produces three collision integrals ${\cal G}^{12}$, ${\cal G}^{22}$ and ${\cal G}^{31}$. Only ${\cal G}^{12}$ and ${\cal G}^{22}$ have been considered by previous authors. The third collision integral ${\cal G}^{31}$ contains the effects of processes where one bogolon becomes three and vice versa. These processes are allowed because the total number of bogolons is not conserved. Since ${\cal G}^{31}$ is of the same order in the interaction strength as ${\cal G}^{22}$, we predict that it will significantly influence the dynamics of the bogolon gas, especially the relaxation of the total number of bogolons to its equilibrium value.
\end{abstract}



%
%
\section{Introduction \label{sec:intro}}

Kinetic equations are a primary tool in the theoretical description of non-uniform gases. They accurately describe the dynamics of dilute gases and are of theoretical interest because they provide a link between microscopic properties and macroscopic observables. A kinetic equation describes the evolution of the relevant distribution function for the system. Kinetic equations generally account for the effects of collisions by including terms that are integrals over the distribution function. These terms are known as collision integrals and provide important information about the behavior of the gas. The structure of the collision integral determines the conserved quantities that are used to parameterize the equilibrium distribution and construct hydrodynamic balance equations. Collision integrals also determine the rate of dissipation and relaxation and can be used to obtain expressions for the transport coefficients.

Kinetic equations have a long history, beginning with the classical Boltzmann equation, which still provides an adequate description of dilute monoatomic gases. Attempts to extend the Boltzmann equation,
 to treat the increasing effects of quantum degeneracy as temperature is lowered,
 resulted in the Uehling-Uhlenbeck (UU) equation \cite{nord, ueh}. However, this equation fails to correctly describe a Bose gas at all temperatures because its steady-state solution is a Bose-Einstein distribution in particle energies. We know that below the critical temperature, many-body effects modify the equilibrium distribution so that it depends on quasiparticle energies rather than particle energies. Many-body effects are accounted for by mean fields that break the $U(1)$ gauge symmetry of the unperturbed Hamiltonian and necessitate a description in terms of quasiparticles. The quasiparticle description is obtained via a Bogoliubov transformation \cite{bogoliubov1}, so we refer to these quasiparticles as bogolons.

Mean fields play a key role in the description of a condensed Bose gas, but determining consistent values for the mean fields is not straightforward. This problem has been discussed by many authors \cite{shi,dodd,fet2,bur,morgan,pet,fet}. Some of the associated issues may be solved by incorporating non-uniformity \cite{griffin2,fedichev} or working in the canonical ensemble as done by V.V. and Vl.V. Kocharovsky \cite{koch,koch2}. However, we note that the structure of the collision integrals only depends on the fact that the $U(1)$ symmetry is broken and does not depend on the particular method of determining the mean fields.

The first attempts \cite{hohen} to derive a kinetic equation for bosons that is valid at all temperatures were motivated by the desire to relate the equations of superfluid hydrodynamics to a microscopic model of interacting bosons. These works did not derive the collision integrals, since they are not necessary to understand the gross features of superfluids. Collision integrals valid below the critical temperature were first presented by Kirkpatrick \& Dorfmann \cite{dorf1}, and were derived by the method of non-equilibrium Green's functions \cite{kada}. They discussed two collision integrals, which we will call ${\cal G}^{12}$ and ${\cal G}^{22}$ that involve  collisions between the bogolons rather than particles.
More recently, Griffin and coworkers \cite{griffin2,hutch,griffin5} have used the non-equilibrium Greens function method to  derive the kinetic equations of a non-uniform condensed Bose gas. Their approach is appropriate for describing the experimentally accessible Bose-Einstein condensates that are produced in trapping potentials. Still, the collision integrals that they derive are the same as those obtained by Kirkpatrick \& Dorfmann.

In this report, we shall derive the collision integrals for a condensed Bose gas using the alternative approach of Peletminskii \& Yatsenko \cite{pel1,pel2}. This approach to kinetic theory has been used by a number of authors to describe relaxation processes in superfluids \cite{galaiko, shum, reichl1, walser}, although none of these deal specifically with dilute BECs. For simplicity, we shall assume that the system is spatially uniform. The methodical nature of this approach allows us to implement the derivation on a computer, which significantly reduces the time required to obtain the collision integrals. In addition to ${\cal G}^{12}$ and ${\cal G}^{22}$, we obtain a third collision integral (${\cal G}^{31}$) that has not been previously discussed.

The collision integral ${\cal G}^{31}$ describes a process in which one bogolon decays into three and vice versa. This process is important to the relaxation of the total bogolon number and is allowed because the number of bogolons is not conserved. In fact, we show that only ${\cal G}^{22}$ conserves bogolon number. All three collision integrals still conserve total energy and momentum. We also show that though the number of bogolons relaxes, the total average number of particles remains constant. The nonconservation of bogolon number means that the BEC has an additional decay mode associated with the relaxation of bogolon number to an equilibrium value. Since previous works do not include ${\cal G}^{31}$, only the effects of ${\cal G}^{12}$ on the relaxation of total bogolon number have been considered. We believe that ${\cal G}^{31}$ plays a dominant role in the relaxation of bogolon number and that it makes a significant contribution to other relaxation processes.

We begin in Sec. \ref{sec:theory} by deriving the mean-field kinetic equation using the PY method. In Sec. \ref{sec:MFham} we introduce the mean field Hamiltonian for a condensed Bose gas and show how can be diagonalized using the Bogoliubov transformation. This diagonalized Hamiltonian governs the dynamics of a gas of bogolons. In Sec. \ref{sec:kineq} we give the collision integrals in the bogolon kinetic equation and in Sec. \ref{sec:prop} we discuss some of their properties. We end in Sec. \ref{sec:conc} with a summary of our results and concluding remarks.

\section{Particle Kinetic Equation \label{sec:theory}}

The basis of the PY method of deriving kinetic equations is the Bogoliubov assumption \cite{bogoliubov2} that for a system that is out of equilibrium, the relaxation to equilibrium can occur in several stages, where each successive stage has a smaller set of relevant parameters (expectation values and mean fields) describing the evolution. The density matrix is then expressed self-consistently as a function of the relevant expectation values and mean fields and the quantum Liouville equation is solved perturbatively. In this approach, setting terms of first order in the interaction to zero defines the mean fields, and terms of second order in the interaction give rise to the collision integrals.

We consider a spatially uniform system of bosons of mass $m$ that are confined to a rectangular box of volume $V$ with periodic boundary conditions. We assume that the particles interact via a contact potential $V({\bf r}_i, {\bf r}_j) = g \delta^3({\bf r}_i - {\bf r}_j)$, where ${\bf r}_i$ is the displacement of the $i^{\rm th}$ particle and $g$ is the strength of the interaction.

The Hamiltonian for this boson gas is
\begin{equation} \label{eq:Ham0}
\hat{H} = \sum_{{\bf k}_1} \epsilon_{{\bf k}_1} \Ad_{{\bf k}_1} \A_{{\bf k}_1} + \frac{g}{2 V} \sum_{{\bf k}_1} \sum_{{\bf k}_2} \sum_{{\bf k}_3} \sum_{{\bf k}_4} \delta_{{\bf k}_1 + {\bf k}_2, {\bf k}_3 - {\bf k}_4} \Ad_{{\bf k}_1} \Ad_{{\bf k}_2} \A_{{\bf k}_3} \A_{{\bf k}_4},
\end{equation}
where $\epsilon_{{\bf k}_1} = \frac{{\hbar}^2 k_1^2}{2 m}$, $\Ad_{{\bf k}_1}$ creates a particle with momentum $\hbar {\bf k}_1$ and $\A_{{\bf k}_1}$ destroys a particle with momentum $\hbar {\bf k}_1$. The creation and annihilation operators satisfy the boson commutation relations $[\A_{{\bf k}_1}, \Ad_{{\bf k}_2}] = \delta_{{\bf k}_1, {\bf k}_2}$ where $\delta_{{\bf k}_1, {\bf k}_2}$ is the product of three Kronecker delta functions, one for each component of ${\bf k}$. The summations run over all single particle states for both positive and negative components of ${\bf k}$. To simplify notation in subsequent sections, we will let $\sum_{{\bf k}_1} \rightarrow \sum_1$, $\Ad_{{\bf k}_1} \rightarrow \Ad_1$, and $\A_{{\bf k}_1} \rightarrow \A_1$. We will keep this subscript convention for all quantities which are function of the wavevector ${\bf k}_i$, such as $\epsilon_1 = \epsilon_{{\bf k}_1}$ and $\delta_{1,2} = \delta_{{\bf k}_1,{\bf k}_2}$.

The full state of the system is described by the full density matrix $\dm(t)$ which obeys the Liouville equation
\begin{equation} \label{eq:liou0}
i \hbar \frac{d \dm}{d t} = [\hat{H}, \dm].
\end{equation}
We now implement the Bogoliubov assumption that, after a short time, the system evolution will relax to one governed by the behavior of the single particle reduced density function and the density matrix $\dm(t)$ will be a functional of the single particle reduced density function
\begin{equation} \label{eq:Gamma}
\Gamma_{i,j}(t) = {\rm Tr}[\dm(t) \hat{\gamma}_{i,j}].
\end{equation}
where
\begin{equation} \label{eq:gamma}
\hat{\gamma}_{i,j} = \left( \begin{array}{cc}
       \Ad_i \A_j   & \Ad_i \Ad_{-j} \\
       \A_{-i} \A_j & \A_{-i} \Ad_{-j}
\end{array}\right).
\end{equation}
Because we consider a spatially homogeneous system, we only need to consider diagonal elements ${\hat \gamma}_{i,i}$ of the more general operator ${\hat \gamma}_{i,j}$. After a sufficiently long time, the density operator can be written
\begin{equation} \label{eq:dm}
\dm(t) = \dm({\bf \Gamma}(t)),
\end{equation}
where ${\bf \Gamma}(t)$ denotes a vector containing $\Gamma_{i,i}(t)$ for all values of $i$. The components   $\Gamma_{i,i}(t)$ are defined self-consistently such that
\begin{equation} \label{eq:Gammaavg}
\Gamma_{i,i}(t) = {\rm Tr}[\dm({\bf \Gamma}(t)) \hat{\gamma}_{i,i}] = \left(\begin{array}{cc}
   \langle \Ad_i \A_i  \rangle  & \langle \Ad_i \Ad_{-i} \rangle \\
   \langle \A_{-i} \A_i \rangle & \langle \A_{-i} \Ad_{-i} \rangle
\end{array}\right)
\end{equation}
The Liouville equation (\ref{eq:liou0}) then takes the form
\begin{equation} \label{eq:liou1}
i \hbar \frac{\partial \dm({\bf \Gamma}(t))}{\partial t} = i \hbar \sum_i \frac{\partial \dm({\bf \Gamma}(t))}{\partial \Gamma^{\mu \nu}_{i,i}(t)} \frac{\partial \Gamma^{\mu \nu}_{i,i}(t)}{\partial t} = [\hat{H}, \dm({\bf \Gamma}(t))],
\end{equation}
where ${\Gamma}^{\mu \nu}_{i,i}(t)$ denotes the $(\mu,\nu)^{\rm th}$ matrix element of the $2 \times 2$ matrix $\Gamma_{i,i}(t)$ and the expression is summed over $\mu$ and $\nu$. The equation for the single particle reduced probability density takes the form
\begin{equation} \label{eq:dGammadt}
i \hbar \frac{\partial \Gamma_{i,i}(t)}{\partial t} = {\rm Tr}\left( \dm({\bf \Gamma}(t)) [ \hat{\gamma}_{i,i}, \hat{H}] \right).
\end{equation}
Let us now combine Eqs. (\ref{eq:liou1}) and (\ref{eq:dGammadt}) to obtain a self-consistent equation for  $\dm({\bf \Gamma}(t))$,
\begin{equation} \label{eq:liou2}
\sum_i \frac{ \partial \dm({\bf \Gamma}(t))}{\partial \Gamma^{\mu \nu}_{i,i}(t)} {\rm Tr} \left( \dm({\bf \Gamma}(t)) [\hat{\gamma}_{i,i}, \hat{H}] \right)^{\mu \nu} = [\hat{H}, \dm({\bf \Gamma}(t))].
\end{equation}
Eq. (\ref{eq:liou2}) is the starting point of the derivation of the  particle kinetic equation. In order to simplify notation for the remainder of this section, we will suppress the dependence of ${\Gamma}_{i,i}(t)$ on time $t$.

The first step in solving Eq. (\ref{eq:liou2}) is the decomposition of the Hamiltonian ${\hat H}$ into a mean field Hamiltonian $\hat{H}^0({\bf \Gamma})$ and an interaction term $\hat{H}^1({\bf \Gamma})$ so that $\hat{H} = \hat{H}^0({\bf \Gamma}) + \hat{H}^1({\bf \Gamma})$. The form of the mean field Hamiltonian $\hat{H}^0({\bf \Gamma})$ is determined by the microscopic conservation laws and broken symmetries and will be written explicitly for BECs in the next section. It phase mixes components $\hat{\gamma}_{i,i}$ of the single particle reduced density function such that
\begin{equation}
[\hat{\gamma}_{i,i}, \hat{H}^0({\bf \Gamma})] = \sum_{j} c_{i,j}({\bf \Gamma}) \hat{\gamma}_{j,j}.
\end{equation}

We can now rewrite Eq. (\ref{eq:liou2}) in terms of $\hat{H}^0({\bf \Gamma})$ and $\hat{H}^1({\bf \Gamma})$. It takes the form
\begin{equation} \label{eq:liou3}
\sum_i \frac{ \partial \dm({\bf \Gamma})}{\partial \Gamma^{\mu \nu}_{i,i}} {\rm Tr} \left( \dm({\bf \Gamma}) [\hat{\gamma}_{i,i}, \hat{H}^0({\bf \Gamma})] \right)^{\mu \nu} - [ \hat{H}^0({\bf \Gamma}), \dm({\bf \Gamma})] = \hat{\mathfrak{F}}({\bf \Gamma})
\end{equation}
where
\begin{equation} \label{eq:frakF}
\hat{\mathfrak{F}}({\bf \Gamma}) = -\sum_i \frac{\partial \dm({\bf \Gamma})}{\partial \Gamma^{\mu \nu}_{i,i}}{\rm Tr} \left( \dm({\bf \Gamma}) [ \hat{\gamma}_{i,i}, \hat{H}^1({\bf \Gamma})] \right)^{\mu \nu} + [ \hat{H}^1({\bf \Gamma}), \dm({\bf \Gamma})].
\end{equation}
We can now solve Eq. (\ref{eq:liou3}) in terms of a perturbation expansion in the interaction $\hat{H}^1({\bf \Gamma})$.

Let us introduce an evolution in fictitious ``time" $s$ governed by the mean field Hamiltonian $\hat{H}^0({\bf \Gamma})$. The $s$-evolution of $\Gamma_{i,i}$ is given by the equation
\begin{equation} \label{eq:dGammads}
i \hbar \frac{\partial \Gamma_{i,i}(s)}{\partial s} \equiv {\rm Tr} \left( \dm({\bf \Gamma}, s) [ \hat{\gamma}_{i,i}, \hat{H}^0({\bf \Gamma})] \right).
\end{equation}
Because $\hat{H}^0({\bf \Gamma})$ commutes with itself, it is independent of $s$. Eq. (\ref{eq:liou3}) can now be written
\begin{equation} \label{eq:liou4}
i \hbar \sum_i \frac{\partial \dm({\bf \Gamma},s)}{\partial \Gamma^{\mu \nu}_{i,i}(s)} \frac{\partial \Gamma^{\mu \nu}_{i,i}(s)}{\partial s} - [\hat{H}^0({\bf \Gamma}), \dm({\bf \Gamma},s)] = \hat{\mathfrak F}({\bf \Gamma},s).
\end{equation}
or,  more simply,
\begin{equation} \label{eq:liou5}
i \hbar \frac{\partial \dm({\bf \Gamma},s)}{\partial s} - [\hat{H}^0({\bf \Gamma}), \dm({\bf \Gamma},s)] = \hat{\mathfrak F}({\bf \Gamma}, s).
\end{equation}
The solution $\dm({\bf \Gamma}, s)$ of Eq. (\ref{eq:liou5}) is also the solution of Eq. (\ref{eq:liou3}) for the case $s=0$.

Using standard methods \cite{pel1}, we can solve Eq. (\ref{eq:liou5}) in terms of a perturbation expansion in the interaction $\hat{H}^1({\bf \Gamma})$. We shall assume that, after very long ``time" $s$, phase mixing induced by $\hat{H}^0({\bf \Gamma})$ and relaxation due to $\hat{H}^1({\bf \Gamma})$ cause the density operator to approach the limiting form
\begin{equation} \label{eq:bound}
\dm_0({\bf \Gamma}) = \lim_{s \to \infty} \hat{U}^0(s, 0) \dm({\bf \Gamma}) \hat{U}^{0 \dagger}(s, 0) = \lim_{s \to -\infty} \hat{U}^0(0, s) \dm({\bf \Gamma}) \hat{U}^{0 \dagger}(0, s)
\end{equation}
where ${\hat U}^0(s_2, s_1) = e^{-i \hat{H}^0({\bf \Gamma}(t))(s_2 - s_1) / {\hbar}}$. In anticipation of the fact that the system will relax to equilibrium, we assume that $\dm_0({\bf \Gamma})$ can be written in the form
\begin{equation} \label{eq:dm00}
\dm_0({\bf \Gamma}) = \exp \left[ -\sum_i X_i \hat{\gamma}_i - \Omega \right].
\end{equation}
where $\Omega = \log \left( {\rm Tr} \left[\exp \left( -\sum_i X_i \hat{\gamma}_i \right) \right] \right)$. The quantities $X_i$ are matrices that encode the values of $\Gamma_{i,i}$. Then the solution to equation (\ref{eq:liou5}) at ``time" $s = 0$ (and therefore also the solution to Eq. (\ref{eq:liou2})) can be written
\begin{equation} \label{eq:rhosoln}
\dm({\bf \Gamma}) = \dm_0({\bf \Gamma}) + \frac{1}{i \hbar} \int_{-\infty}^0 d s \hat{U}^0 (0, s) \hat{\mathfrak F}({\bf \Gamma}, s) \hat{U}^{0 \dagger}(0, s).
\end{equation}
The fact that ${\rm Tr} \left(\hat{\mathfrak F}({\bf \Gamma}) \hat{\gamma}_{j,j} \right) = 0$ implies that ${\rm Tr} \left(\dm({\bf \Gamma}) \hat{\gamma}_{j,j}\right) = {\rm Tr} \left(\dm_0({\bf \Gamma}) \hat{\gamma}_{j,j}\right)$. This means that we can evaluate $\Gamma_{i,i}$ by using the known density matrix $\hat{\rho}_0$. To show that ${\rm Tr} \left(\hat{\mathfrak F}({\bf \Gamma}) \hat{\gamma}_{j,j} \right) = 0$, notice that when we take the trace of the first term of Eq. (\ref{eq:frakF}) after it has been multiplied by $\hat{\gamma}_{j,j}$ we obtain $\frac{\partial \Gamma_{i,i}}{\partial \Gamma_{j,j}} = \delta_{i,j}$ which eliminates the summation. The two remaining terms in sum Eq. (\ref{eq:frakF}) to zero.

We now expand this solution to second order in $\hat{H}^1({\bf \Gamma})$. One can consider higher order terms, but the expansion to second order is sufficient for the case of a dilute BEC. The mean field Hamiltonian is defined so the first order contribution is zero,
\begin{equation} \label{eq:MFcond}
{\rm Tr}\left(\dm_0({\bf \Gamma}) [\hat{\gamma}_{i,i}, \hat{H}^1({\bf \Gamma})] \right) = 0.
\end{equation}
This eliminates first order terms from the kinetic equation. To second order in the interaction, Eq. (\ref{eq:dGammadt}) takes the form
\begin{equation} \label{eq:kineq0}
\begin{split}
\frac{\partial \Gamma_{i,i}(t)}{\partial t} = \frac{1}{i \hbar} \sum_{j} c_{i,j}({\bf \Gamma}) \Gamma_{j,j} + \frac{1}{\hbar^2} \int_{-\infty}^0 d s {\rm Tr} \left( \dm_0({\bf \Gamma}) [\hat{H}^1({\bf \Gamma}), \hat{U}^{0 \dagger}(0, s) [\hat{\gamma}_{i,i}, \hat{H}^1({\bf \Gamma})] \hat{U}^0(0, s) ] \right).
\end{split}
\end{equation}
This is the kinetic equation that is generated by the PY method. In the next section, we choose a form of $\hat{H}^0({\bf \Gamma})$ that makes the first term vanish. The second term on the right hand side gives rise to the collision integrals. To evaluate it, we must now provide explicit the forms of $\hat{H}^0({\bf \Gamma})$ and $\hat{H}^1({\bf \Gamma})$ that are appropriate to our system.

\section{Mean Field Hamiltonian and Bogoliubov Transformation
\label{sec:MFham}}

We can write the Hamiltonian for the BEC in a way that allows for the broken gauge symmetry and conserves the {\it average}  number of particles $N$. We also want the form of $\hat{H}^1$ to be such that the condition (\ref{eq:MFcond}) can be satisfied. A mean field Hamiltonian $\hat{H} = \hat{H}^0 + \hat{H}^1$ that satisfies all of these conditions can be written in general as
\begin{equation} \label{eq:MFH0}
\hat{H}^0 = \Xi + {\sum_i} \left[ (\epsilon_i - \mu + \nu) \Ad_i \A_i + \frac{\Delta}{2} (\Ad_i \Ad_{-i} + \A_{-i} \A_i) \right],
\end{equation}
and
\begin{equation} \label{eq:H1}
\hat{H}^1 = \frac{g}{2 V} {\sum_{ijkl}} \delta_{i+j, k+l} \Ad_i \Ad_j \A_k \A_l - {\sum_i} \left[ \nu \Ad_i \A_i + \frac{\Delta}{2} (\Ad_i \Ad_{-i} + \A_{-i} \A_i) \right] - \Xi,
\end{equation}
where the quantities $\nu$ and $\Delta$ are mean fields given by
\begin{equation} \label{eq:nueqAA}
\nu = \frac{2 g}{V} {\sum_i} {\rm Tr} (\dm_0 \Ad_i \A_i) = \frac{2 g N}{V},
\end{equation}
and
\begin{equation} \label{eq:deeqAA}
\Delta = \frac{g}{V} {\sum_i} {\rm Tr} (\dm_0 \Ad_i \Ad_{-i}) = \frac{g}{V} {\sum_i} {\rm Tr} (\dm_0 \A_{-i} \A_i),
\end{equation}
and $\mu$ is the chemical potential. Note that though the average number of particles is conserved by the total Hamiltonian $\hat{H}$, it is not conserved by $\hat{H}^0$. The quantity $\Xi$ is a shift in the energy of the system and is chosen so that ${\rm Tr} \left( \dm_0 \hat{H}^1 \right) = 0$ at zero temperature. Since $\Xi$ is not an operator, it has no effect on our derivation of the kinetic equation.

The mean field Hamiltonian $\hat{H}^0$ can be diagonalized by a Bogoliubov transformation. We define the operators
\begin{equation} \label{eq:Bd}
\Bd_i = u_i \Ad_i + v_i \A_{-i} \hspace{0.5in} {\rm and} \hspace{0.5in} \B_i = u_i \A_i + v_i \Ad_{-i}
\end{equation}
with
\begin{equation}
u_i = \frac{1}{\sqrt{2}}\sqrt{\frac{\epsilon_i + \nu - \mu}{E_i} + 1}, \hspace{1in} v_i = \frac{1}{\sqrt{2}}\sqrt{\frac{\epsilon_i + \nu - \mu}{E_i} - 1}
\end{equation}
where
\begin{equation} \label{eq:bogenergy}
E_i = \sqrt{(\epsilon_i + \nu - \mu)^2 - \Delta^2}.
\end{equation}
The operators $\Bd_i$ and $\B_j$ have the property that $[\B_i, \Bd_j] = \delta_{i,j}$ and in terms of them, the mean field Hamiltonian is
\begin{equation} \label{eq:boghamil}
\hat{H}^0 = {\sum_i} E_i \Bd_i \B_i + \frac{1}{2}{\sum_i} \left( E_i - \epsilon_i - \nu + \mu \right) + \Xi.
\end{equation}
Therefore, $\Bd_i$ and $\B_i$ are interpreted as creation and annihilation operators for excitations, which we will refer to as bogolons.

It is well known that the energy spectrum of bosonic excitations must be gapless, that is, $E_i$ must approach zero as $i \to 0$.  This implies that $\mu = \nu - \Delta$. In the Green's function approach, this statement is known as the Hugenholtz-Pines (HP) theorem \cite{hug} and is expressed in terms of the self-energies as $\Sigma_{11} - \Sigma_{12} = \mu$. The self energies are not known exactly, but can be calculated at several different levels of approximation \cite{griffin5}. It is apparent that our mean fields will correspond to the self energies evaluated within one of these approximations. Setting $\mu = \nu - \Delta$ requires us to give special treatment to the $i = 0$ operators, since $u_i$ and $v_i$ are undefined (become infinite) when $E_i = 0$. This is to be expected in a BEC because the zero momentum state actually is special; it contains the condensate. Following Bogoliubov, we implement the special treatment of $i = 0$ operators by replacing $\Ad_0$ and $\A_0$ with $\sqrt{N_0}$ where $N_0$ is the number of particles in the $i = 0$ state.

We can now evaluate the expectation values appearing in the self-consistency equations (\ref{eq:nueqAA}) and (\ref{eq:deeqAA})
as well as the collision integrals. Since the unperturbed Hamiltonian $\hat{H}^0$ is diagonal with respect to the operators $\Bd_i \B_i$, the density matrix $\dm_0$ has a Gaussian form. This allows us to use Wick theorem to evaluate higher-order expectation values of bogolon creation and annihilation operators. We also note that whenever zero appears in a particle operator momentum index, $\Ad_0$ and $\A_0$ must be replaced with $\sqrt{N_0}$ before expectation values are taken.

Using the Bogoliubov replacement and the Bogoliubov transformation, we find that the self-consistency equations become
\begin{equation} \label{eq:nueqBB}
\nu = \frac{2 g N_0}{V} + \frac{2 g}{V}{\sum_i}' (u_i^2 + v_i^2) \langle \Bd_i \B_i \rangle + \frac{2 g}{V}{\sum_i}' v_i^2
\end{equation}
and
\begin{equation} \label{eq:deeqBB}
\Delta = \frac{g N_0}{V} - \frac{2 g}{V} {\sum_i}' u_i v_i \langle \Bd_i \B_i \rangle - \frac{g}{V} {\sum_i}' u_i v_i.
\end{equation}
These equations are identical to those obtained for the self-energies $\Sigma_{11}$ and $\Sigma_{12}$ evaluated in the Hartree-Fock-Bogoliubov (HFB) approximation \cite{bur,fet,lee}. It is well known that the self energies of the HFB approximation do not obey the Hugenholtz-Pines theorem. Following the lead of previous authors for  dilute BECs \cite{koch,koch2}, we can choose to keep the HP theorem valid by using the Popov approximation to the self-consistency equations. In the Popov approximation, $\Delta$ is set equal to $\frac{g N_0}{V}$. For dilute BECs, this approximation has been shown to give good agreement with experiment as long as the temperature is low compared to the critical temperature $T_C$, namely $T < 0.6 T_C$. Much work has been done on more general  approximations of the self-energies \cite{morgan,griffin5,popov1,popov2,beliaev} and more sophisticated approximations that work well at higher temperatures are available.   It is important to note, however,  that the derivation of the collision integrals does not depend on the method used to obtain the mean field values.Ó

\section{Derivation of Collision Integrals \label{sec:kineq}}

In this section, we turn our attention towards calculation of the collision integrals. The bogolon expectation values $\langle \Bd_i \B_i \rangle$ along with $N_0$ will form a closed set of kinetic variables. In order to evaluate the kinetic equations for these variables, we first note that the time derivative of the bogolon expectation values can be written
\begin{equation} \label{eq:BdB}
\frac{d \langle \Bd_i \B_i \rangle}{d t} = (u_i^2 + v_i^2) \frac{d \langle \Ad_i \A_i \rangle}{d t} + u_i v_i \frac{d \langle \Ad_i \Ad_{-i} \rangle}{d t} + u_i v_i \frac{d \langle \A_{-i} \A_i \rangle}{d t}.
\end{equation}
The absence of any time derivatives on $u_i$ and $v_i$ is because their time derivatives can be written as $\frac{d u_i}{d t} = v_i \frac{d \theta_i}{d t}$ and $\frac{d v_i}{d t} = u_i \frac{d \theta_i}{d t}$ where $\theta_i = \frac{1}{2} \cosh^{-1} \left( \frac{\epsilon_i + \nu - \mu}{E_i} \right)$. The coefficient of $\frac{d \theta_i}{d t}$ in Eq. \ref{eq:BdB} can be expressed in terms of bogolon occupation numbers and is found to vanish identically.

Let us introduce the notation
\begin{equation}
\N_i \equiv \langle \Bd_i \B_i \rangle \hspace{1in} \F_i \equiv \langle \B_i \Bd_i \rangle = 1 + \N_i,
\end{equation}
\begin{equation}
N_i \equiv \langle \Ad_i \A_i \rangle \hspace{1in} F_i \equiv \langle \A_i \Ad_i \rangle = 1 + N_i
\end{equation}
and
\begin{equation}
\Lambda_i \equiv \frac{1}{2} \left( \langle \Ad_i \Ad_{-i} \rangle + \langle \A_{-i} \A_i \rangle \right).
\end{equation}
The calculation of $\frac{d \N_i}{d t}$ is broken up into the calculation of $\frac{d N_i}{d t}$ and $\frac{d \Lambda_i}{d t}$. This provides us with more information concerning the evolution of the particle occupation numbers $N_i$ than calculating $\frac{d \N_i}{dt}$ directly.

We can now use Eq. (\ref{eq:H1}) and the bogolon operators to evaluate the right hand side of Eq. (\ref{eq:kineq0}). To accomplish this, all particle operators except $\Ad_0$ and $\A_0$ are expressed in terms of bogolon operators and the $s$-evolution of the bogolon operators is resolved. Commutators are then evaluated and the remaining operators are put in normal order. Finally, the expectation values are evaluated using Wick expansions. This process generates a very large number of individual terms. The labor of condensing these terms to a manageable size was greatly reduced through the use of a custom computer algebra code that we developed.

The kinetic equation for the expectation value $N_i$ (the particle distribution) can be condensed to the form
\begin{equation} \label{eq:particlekineq}
\frac{d N_i}{d t} = \mathcal{C}^{12}_i \{\N\} + \mathcal{C}^{22}_i \{\N\} + \mathcal{C}^{31}_i \{\N\},
\end{equation}
where $\mathcal{C}^{12}_i$, $\mathcal{C}^{22}_i$ and $\mathcal{C}^{31}_i$ are collision integrals given explicitly in appendix \ref{ax:colop}. The kinetic equation for the value $\Lambda_i \equiv \frac{1}{2} \left( \Ad_i \Ad_{-i} + \A_{-i} \A_i \right)$ has a similar structure but with different collision integrals. It can be written as
\begin{equation} \label{eq:anomkineq}
\frac{d \Lambda_i}{d t} = \mathcal{D}^{12}_i \{\N\} + \mathcal{D}^{22}_i \{\N\} + \mathcal{D}^{31}_i \{\N\},
\end{equation}
where expressions for $\mathcal{D}^{12}_i$, $\mathcal{D}^{22}_i$ and $\mathcal{D}^{31}_i$ can also be found in appendix \ref{ax:colop}.

To obtain an equation for the evolution of $\N_i$ (the bogolon distribution) , we use Eq. (\ref{eq:BdB}), (\ref{eq:particlekineq}) and (\ref{eq:anomkineq}). Putting these together results in a closed kinetic equation for the bogolon distribution $\N_i$,
\begin{equation} \label{eq:bogokineq}
\frac{d \N_i}{d t} = \mathcal{G}^{12}_i \{\N\} + \mathcal{G}^{22}_i \{\N\} + \mathcal{G}^{31}_i \{\N\}
\end{equation}
where the bogolon collision integrals are given by
\begin{equation} \label{eq:G12}
\begin{split}
\mathcal{G}^{12}_i \{\N\} = \frac{8 \pi N_0 g^2}{\hbar V^2} &{\sum_{2,3}}' \delta_{i+2,3} \delta(E_i + E_2 - E_3) (W^{12}_{i,2,3})^2 (\F_i \F_2 \N_3 -
\N_i \N_2 \F_3) \\
+ \frac{4 \pi N_0 g^2}{\hbar V^2} &{\sum_{2,3}}' \delta_{i,2+3} \delta(E_i - E_2 - E_3) (W^{12}_{3,2,i})^2 (\F_i \N_2 \N_3 - \N_i \F_2 \F_3),
\end{split}
\end{equation}
\begin{equation} \label{eq:G22}
\mathcal{G}^{22}_i \{\N\} = \frac{4 \pi g^2}{\hbar V^2} {\sum_{2,3,4}}' \delta_{i+2,3+4} \delta(E_i + E_2 - E_3 - E_4) (W^{22}_{i,2,3,4})^2 (\F_i \F_2 \N_3 \N_4 - \N_i \N_2 \F_3 \F_4)
\end{equation}
and
\begin{equation} \label{eq:G31}
\begin{split}
\mathcal{G}^{31}_i \{\N\} = \frac{4 \pi g^2}{3 \hbar V^2} &{\sum_{2,3,4}}' \delta_{i,2+3+4} \delta(E_i - E_2 - E_3 - E_4) (W^{31}_{1,2,3,4})^2 (\F_i \N_2
\N_3 \N_4 - \N_i \F_2 \F_3 \F_4) \\
+ \frac{4 \pi g^2}{\hbar V^2} &{\sum_{2,3,4}}' \delta_{i+2+3,4} \delta(E_i + E_2 + E_3 - E_4) (W^{31}_{4,3,2,i})^2 (\F_i \F_2 \F_3 \N_4 - \N_i \N_2 \N_3 \F_4).
\end{split}
\end{equation}

The weighting functions are given in terms of $u_i$ and $v_i$ by
\begin{equation}
W^{12}_{1,2,3} = u_1 u_2 u_3 - u_1 v_2 u_3 - v_1 u_2 u_3 + u_1 v_2 v_3 + v_1 u_2 v_3 - v_1 v_2 v_3,
\end{equation}
\begin{equation}
W^{22}_{1,2,3,4} = u_1 u_2 u_3 u_4 + u_1 v_2 u_3 v_4 + u_1 v_2 v_3 u_4 + v_1 u_2 u_3 v_4 + v_1 u_2 v_3 u_4 + v_1 v_2 v_3 v_4
\end{equation}
and
\begin{equation}
W^{31}_{1,2,3,4} = u_1 u_2 u_3 v_4 + u_1 u_2 v_3 u_4 + u_1 v_2 u_3 u_4 + v_1 v_2 v_3 u_4 + v_1 v_2 u_3 v_4 + v_1 u_2 v_3 v_4.
\end{equation}
Each of these weighting functions has specific symmetry with respect to interchanges of its indices that is shared by its collision integral. The collision integrals $\mathcal{G}^{12}$ and ${\cal G}^{22}$ are identical to those generated by the Green's function approach\cite{dorf1}. The collision integral $\mathcal{G}^{31}$ appears as a natural result of our calculation using the PY approach.

If one sets $\Delta = 0$, the weighting functions $W^{12}_{i,j,k}$ and $W^{22}_{i,j,k,l}$ equal $1$ while the weighting function $W^{31}_{i,j,k,l}$ vanishes. The self-consistency equations with $\Delta = 0$ imply that $N_0$ also equals zero. What remains of the kinetic equation when $\Delta = 0$ is just the Uehling-Uhlenbeck equation. Therefore, the kinetic equation appropriate for an ideal Bose gas is recovered above the critical temperature.

The three basic processes available to bogolons are represented by the three collision integrals. In ${\cal G}^{12}$, one  bogolon  decays into two bogolons. In ${\cal G}^{22}$, two bogolons collide elastically with each other. The collision integral ${\cal G}^{31}$ describes one bogolon decaying into three and its inverse process.

\section{Properties of the Collision Integrals \label{sec:prop}}

The kinetic equation conserves the total bogolon energy given by ${\sum_i}' E_i \N_i$ and total bogolon momentum given by ${\sum_i}' i \N_i$. In fact, each collision integral conserves energy and momentum separately. This can by shown by evaluating the sums
\begin{equation}
{\sum_i}' E_i {\cal G}^{12}_i \{\N\} = {\sum_i}' E_i {\cal G}^{22}_i \{\N\} = {\sum_i}' E_i {\cal G}^{31}_i \{\N\} = 0
\end{equation}
and
\begin{eqnarray}
{\sum_i}' i {\cal G}^{12}_i \{\N\} = {\sum_i}' i {\cal G}^{22}_i \{\N\} = {\sum_i}' i {\cal G}^{31}_i \{\N\} = 0.
\end{eqnarray}
This calculation relies on the symmetry properties of the collision integral and the weighting functions to rename summation indices.

One can show that the total bogolon number given by ${\sum_i}' \N_i$ is not conserved. This follows from the fact that
\begin{equation}
{\sum_i}' {\cal G}^{12}_i \{\N\} \neq 0 \hspace{0.5in} {\rm and} \hspace{0.5in} {\sum_i}' {\cal G}^{31}_i \{\N\} \neq 0.
\end{equation}
However, it can still be shown that
\begin{equation}
{\sum_i}' {\cal G}^{22}_i \{\N\} = 0
\end{equation}
meaning that ${\cal G}^{22}$ acting alone conserves bogolon number, as we would expect from a two-body elastic collision.

In each of the three collision processes, the rate of the process depends upon the bogolon momentum distribution $\N_i$ through the factors such as $\F_i \F_2 \N_3 \N_4 - \N_i \N_2 \F_3 \F_4$. The evolution of the distribution function proceeds until the rate of each process is balanced by the rate of its inverse process and equilibrium is achieved. With generality, we only consider equilibrium states of zero total momentum. A quick analysis of the collision integrals shows that equilibrium distribution $\N^0_i$ must be a Bose-Einstein distribution in the bogolon energies.
\begin{equation}
\N^0_i = \frac{1}{e^{E_i / (k_B T)} - 1}
\end{equation}
The bogolons have zero chemical potential since the total number of bogolons is not conserved. The effects of the particle's chemical potential are included through the bogolon energy spectrum $E_i$ given by Eq. (\ref{eq:bogenergy}). It is easy to verify that $\N^0_i$ is a steady state solution of the kinetic equation by substitution into Eq. (\ref{eq:bogokineq}).

It is interesting to use the kinetic equation to evaluate the time rate of change of the total particle number given by
$\frac{d N}{d t} = {\sum_i} \frac{d}{d t} \langle \Ad_i \A_i \rangle$. This sum can be evaluate in two parts as
\begin{equation}
\frac{d N}{d t} = \frac{d N_0}{d t} + {\sum_i}' \frac{d}{d t} \langle \Ad_i \A_i \rangle.
\end{equation}
Using Eq. (\ref{eq:particlekineq}), we can show that
\begin{equation}
{\sum_i}' \frac{d}{d t} \langle \Ad_i \A_i \rangle = - \frac{4 \pi N_0 g^2}{\hbar V^2} {\sum_{1,2,3}}' \delta_{1,2+3} \delta(E_1 - E_2 - E_3)
 W^{12}_{3,2,1} \tilde{W}^{12}_{1,2,3} (\F_1 \N_2 \N_3 - \N_1 \F_2 \F_3)
\end{equation}
where
\begin{equation}
\tilde{W}^{12}_{1,2,3} = u_1 u_2 u_3 + u_1 u_2 v_3 + u_1 v_2 u_3 + v_1 u_2 v_3 + v_1 v_2 u_3 + v_1 v_2 v_3
\end{equation}
As expected, this is non-zero because the ${\cal G}^{12}$ collision integral contains collisions with the condensate. From the fact that the total average number of particles is conserved, we should expect that $\frac{d N_0}{d t} = - {\sum_i}' \frac{d}{d t} \langle \Ad_i \A_i \rangle$. However, to show this from  Eq. (\ref{eq:particlekineq}), we must adopt a convention for the Bogoliubov transformation at zero momentum. In all we have done so far, the Bogoliubov transformation does not act on the zero momentum operators and the particle operators $\Ad_0$ and $\A_0$ are treated separately. This is equivalent to defining $u_0 = 1$ and $v_0 = 0$ and taking $\N_0 = \F_0 = N_0$. Using these substitutions to compute $\frac{d N_0}{d t}$ from Eq. (\ref{eq:particlekineq}), we obtain
\begin{equation}
\frac{d N_0}{d t} = \frac{4 \pi N_0 g^2}{\hbar V^2} {\sum_{1,2,3}}' \delta_{1,2+3} \delta(E_1 - E_2 - E_3) W^{12}_{3,2,1} \tilde{W}^{12}_{1,2,3} (\F_1 \N_2 \N_3 - \N_1 \F_2 \F_3).
\end{equation}
This explicitly shows that $\frac{d N_0}{d t} = - {\sum_i}' \frac{d}{d t} \langle \Ad_i \A_i \rangle$ and therefore the total average particle number is conserved.

\section{Conclusions \label{sec:conc}}

We have derived a kinetic equation for the momentum distribution of Bogoliubov excitations (bogolons) in a spatially uniform condensed Bose gas using the approach of Peletminskii and Yatsenko.  The kinetic equation contains three collision integrals, ${\cal G}^{22}$ , ${\cal G}^{21}$ , and ${\cal G}^{31}$  that describe the bogon processes that lead to decay to equilibrium of the BEC. The kinetic equation conserves total particle number, total energy and total momentum, but does not conserve total bogolon number.

The collision integral ${\cal G}^{31}$ has not been analyzed by previous authors. We can show (to appear elsewhere) that it has a significant influence on dissipation and relaxation processes in the gas. Since ${\cal G}^{31}$ describes a process that does not conserve bogolon number, we find that its most significant influence is on the relaxation of the total bogolon number to its equilibrium value. However, this collision integral also has a measurable influence on other dissipative properties of the Bose-condensed gas, such as relaxation rates and transport coefficients.Ó

\section{Acknowledgements}

The authors wish to thank the Robert A. Welch Foundation (Grant No.  F-1051) for support of this work. The authors thank V.V and Vl.V Kocharovsky for initial discussions that lead to this work.

\appendix
\section{Derivation of Collision Operators \label{ax:colop}}

In this appendix, we give explicit forms for the collision integrals $\mathcal{C}$ and $\mathcal{D}$ found in Eqs. (\ref{eq:particlekineq}) and (\ref{eq:anomkineq}). To generate the kinetic equations for the quantities $N_i$ and $\Lambda_i$, we use Eq. (\ref{eq:kineq0}) to get

\begin{equation} \label{eq:dNidt}
\begin{split}
\frac{d N_i}{d t} =& \frac{g^2}{4 \hbar^2 V^2} \sum_{1,2,3,4} \sum_{5,6,7,8} \delta_{1+2, 3+4} \delta_{5+6, 7+8} \\ &\times \int\limits_{-\infty}^0 d s \langle [\Ad_1 \Ad_2 \A_3 \A_4, [\Ad_i(s) \A_i(s), \Ad_5(s) \Ad_6(s) \A_7(s) \A_8(s)] \rangle
\end{split}
\end{equation}
and
\begin{equation} \label{eq:dLambdaidt}
\begin{split}
\frac{d \Lambda_i}{d t} =& \frac{g^2}{8 \hbar^2 V^2} \sum_{1,2,3,4}
\sum_{5,6,7,8} \delta_{1+2, 3+4} \delta_{5+6, 7+8} \\
&\times \int\limits_{-\infty}^0 d s \langle [\Ad_1 \Ad_2 \A_3 \A_4, [\Ad_i(s)
\Ad_{-i}(s), \Ad_5(s) \Ad_6(s) \A_7(s) \A_8(s)] \rangle \\
+& \frac{g^2}{8 \hbar^2 V^2} \sum_{1,2,3,4} \sum_{5,6,7,8} \delta_{1+2, 3+4}
\delta_{5+6, 7+8} \\
&\times \int\limits_{-\infty}^0 d s \langle [\Ad_1 \Ad_2 \A_3 \A_4, [\A_{-i}(s) \A_i(s), \Ad_5(s) \Ad_6(s) \A_7(s) \A_8(s)] \rangle
\end{split}
\end{equation}
The calculation of the commutations and traces in Eqs. (\ref{eq:dNidt}) and (\ref{eq:dLambdaidt}) is quite arduous. To speed up the process, we developed a custom symbolic algebra code which can store and manipulate large expressions containing both operators and c-number variables. The code is capable of expanding commutators, evaluating expectation values according to Wick factorization rules, performing sums over Kronecker delta functions and permuting summation indices to recombine terms in a compact form. An output file is generated containing an algebraic expression that is anywhere from 70 to 500 terms long. The final simplification and factorization is done by hand. Following this process, we obtain the Eq. (\ref{eq:particlekineq}) with
\begin{equation}
\begin{split}
\mathcal{C}^{12}_1 \{\N\} =& \frac{4 \pi N_0 g^2}{\hbar V^2} {\sum_{2,3}}' \delta_{1,2+3} \delta(E_1 -
E_2 - E_3) W^{12}_{3,2,1} \\
& \times \left[ \Upsilon^A_{1,2,3} (\F_1 \N_2 \N_3 - \N_1 \F_2 \F_3) + \tilde{\Upsilon}^A_{1,2,3} (\F_{-1} \N_{-2} \N_{-3} - \N_{-1} \F_{-2}
\F_{-3}) \right] \\
+& \frac{8 \pi N_0 g^2}{\hbar V^2} {\sum_{2,3}}' \delta_{1+2,3} \delta(E_1 +
E_2 - E_3) W^{12}_{1,2,3} \\
& \times \left[\Upsilon^B_{1,2,3} (\F_1 \F_2 \N_3 - \N_1 \N_2 \F_3) + \tilde{\Upsilon}^B_{1,2,3} (\F_{-1} \F_{-2} \N_{-3} - \N_{-1} \N_{-2} \F_{-3}) \right],
\end{split}
\end{equation}
\begin{equation}
\begin{split}
\mathcal{C}^{22}_1 \{\N\} =& \frac{4 \pi g^2}{\hbar V^2} {\sum_{2,3,4}}' \delta_{1+2,3+4} \delta(E_1 + E_2 - E_3 - E_4) W^{22}_{1,2,3,4} \\
&\times \Big[ \Upsilon^C_{1,2,3,4} (\F_1 \F_2 \N_3 \N_4 - \N_1 \N_2 \F_3 \F_4) \\
& ~~ - \tilde{\Upsilon}^C_{1,2,3,4} (\F_{-1} \F_{-2} \N_{-3} \N_{-4} - \N_{-1} \N_{-2} \F_{-3} \F_{-4}) \Big]
\end{split}
\end{equation}
and
\begin{equation}
\begin{split}
\mathcal{C}^{31}_1 \{\N\} =& \frac{4 \pi g^2}{3 \hbar V^2} {\sum_{2,3,4}}' \delta_{1,2+3+4} \delta(E_1 - E_2 - E_3 - E_4) W^{31}_{1,2,3,4}  \\
& \times \Big[ \Upsilon^D_{1,2,3,4} (\F_1 \N_2 \N_3 \N_4 - \N_1 \F_2 \F_3 \F_4 ) \\
& ~~ - \tilde{\Upsilon}^D_{1,2,3,4} (\F_{-1} \N_{-2} \N_{-3} \N_{-4} - \N_{-1} \F_{-2} \F_{-3} \F_{-4}) \Big] \\
+&\frac{4 \pi g^2}{\hbar V^2} {\sum_{2,3,4}}' \delta_{1+2+3,4} \delta(E_1 +E_2 + E_3 - E_4) W^{31}_{4,3,2,1}  \\
&\times \Big[ \Upsilon^E_{1,2,3,4} (\F_1 \F_2 \F_3 \N_4 - \N_1 \N_2 \N_3 \F_4) \\
& ~~ - \tilde{\Upsilon}^E_{1,2,3,4} (\F_{-1} \F_{-2} \F_{-3} \N_{-4} - \N_{-1} \N_{-2} \N_{-3} \F_{-4}) \Big]
\end{split}
\end{equation}
where
\begin{eqnarray}
\Upsilon^A_{1,2,3} &=& u_1 u_2 u_3 - u_1 v_2 u_3 - u_1 u_2 v_3 \\
\Upsilon^B_{1,2,3} &=& u_1 u_2 u_3 + u_1 v_2 v_3 - u_1 v_2 u_3 \\
\Upsilon^C_{1,2,3,4} &=& u_1 u_2 u_3 u_4 + u_1 v_2 v_3 u_4 + u_1 v_2 u_3 v_4 \\
\Upsilon^D_{1,2,3,4} &=& u_1 v_2 u_3 u_4 + u_1 u_2 v_3 u_4 + u_1 u_2 u_3 v_4 \\
\Upsilon^E_{1,2,3,4} &=& u_1 u_2 v_3 u_4 + u_1 v_2 u_3 u_4 + u_1 v_2 v_3 v_4
\end{eqnarray}
and $\tilde{\Upsilon}$ is $\Upsilon$ with each $u$ and $v$ interchanged.

For $\Lambda_i$ we obtain Eq. (\ref{eq:anomkineq}) with
\begin{equation}
\begin{split}
\mathcal{D}^{12}_1 \{\N\} =& - \frac{4 \pi N_0 g^2}{\hbar V^2} {\sum_{2,3}}'
\delta(E_1 + E_2 - E_3) \delta_{1+2,3} W^{12}_{1,2,3} \Omega^A_{1,2,3} \\
& \times (\F_1 \F_2 \N_3 - \N_1 \N_2 \F_3 + \F_{-1} \F_{-2} \N_{-3} - \N_{-1}
\N_{-2} \F_{-3}) \\
+& \frac{2 \pi N_0 g^2}{\hbar V^2} {\sum_{2,3}}' \delta(E_1 - E_2 - E_3)
\delta_{1,2+3} W^{12}_{3,2,1} \Omega^B_{1,2,3} \\
&\times (\F_1 \N_2 \N_3 - \N_1 \F_2 \F_3 + \F_{-1} \N_{-2} \N_{-3} - \N_{-1} \F_{-2} \F_{-3}),
\end{split}
\end{equation}
\begin{equation}
\begin{split}
\mathcal{D}^{22}_1 \{\N\} =& \frac{2 \pi g^2}{\hbar V^2} {\sum_{2,3,4}}' \delta(E_1 + E_2 - E_3 - E_4) \delta_{1+2,3+4} W^{22}_{1,2,3,4}
\Omega^C_{1,2,3,4} \\
&\times (\F_1 \F_2 \N_3 \N_4 - \N_1 \N_2 \F_3 \F_4 + \F_{-1} \F_{-2} \N_{-3} \N_{-4} - \N_{-1} \N_{-2} \F_{-3} \F_{-4})
\end{split}
\end{equation}
and
\begin{equation}
\begin{split}
\mathcal{D}^{31}_1 \{\N\} =& -\frac{2 \pi g^2}{3 \hbar V^2} {\sum_{2,3,4}}' \delta(E_1 - E_2 - E_3 - E_4) \delta_{1,2+3+4} W^{31}_{1,2,3,4}
\Omega^D_{1,2,3,4} \\
&\times (\F_1 \N_2 \N_3 \N_4 - \N_1 \F_2 \F_3 \F_4 + \F_{-1} \N_{-2} \N_{-3}
\N_{-4} - \N_{-1} \F_{-2} \F_{-3} \F_{-4}) \\
+&\frac{2 \pi g^2}{\hbar V^2} {\sum_{2,3,4}}' \delta(E_1 + E_2 + E_3 -
E_4) \delta_{1+2+3,4} W^{31}_{4,3,2,1} \Omega^E_{1,2,3,4} \\
& \times (\F_1 \F_2 \F_3 \N_4 - \N_1 \N_2 \N_3 \F_4 + \F_{-1} \F_{-2} \F_{-3} \N_{-4} - \N_{-1} \N_{-2} \N_{-3} \F_{-4})
\end{split}
\end{equation}
where
\begin{eqnarray}
\Omega^A_{1,2,3} &=& u_1 v_2 v_3 + v_1 v_2 v_3 - v_1 v_2 u_3 + v_1 u_2 u_3 + u_1
u_2 u_3 - u_1 u_2 v_3 \\
\Omega^B_{1,2,3} &=& v_1 v_2 u_3 - u_1 v_2 v_3 + u_1 u_2 v_3 + u_1 v_2 u_3 + v_1
u_2 v_3 - v_1 u_2 u_3 \\
\Omega^C_{1,2,3,4} &=& u_1 v_2 v_3 v_4 + u_1 u_2 v_3 u_4 + u_1 u_2 u_3 v_4 - v_1
v_2 v_3 u_4 - v_1 u_2 u_3 u_4 - v_1 v_2 u_3 v_4 \\
\Omega^D_{1,2,3,4} &=& v_1 v_2 u_3 u_4 + v_1 u_2 v_3 u_4 + v_1 u_2 u_3 v_4 - u_1
u_2 v_3 v_4 - u_1 v_2 u_3 v_4 - u_1 v_2 v_3 u_4 \\
\Omega^E_{1,2,3,4} &=& u_1 u_2 u_3 u_4 + u_1 v_2 u_3 v_4 + u_1 v_2 u_3 v_4 - v_1 u_2 v_3 u_4 - v_1 u_2 v_3 u_4 - v_1 v_2 v_3 v_4.
\end{eqnarray}
When these two results are combined using Eq. (\ref{eq:BdB}), we obtain Eq. (\ref{eq:bogokineq}).

\end{document}